\documentclass[12pt]{article}

\usepackage[UKenglish]{babel}

\usepackage{latexsym}
\usepackage{amssymb}
\usepackage{epsfig}

\begin{document}

\title{Anomalous localisation in the aperiodic Kronig-Penney model}

\author{ J. C. Hern\'{a}ndez-Herrej\'{o}n${}^{1}$, F.~M.~Izrailev${}^{2}$,
and L.~Tessieri${}^{1}$ \\
{\it ${}^{1}$ Instituto de F\'{\i}sica y Matem\'{a}ticas,} \\
{\it Universidad Michoacana de San Nicol\'{a}s de Hidalgo,} \\
{\it Morelia, Mich., 58060, Mexico} \\
{\it ${}^{2}$ Instituto de F\'{\i}sica, Universidad Aut\'{o}noma de Puebla,} \\
{\it Puebla, Pue., 72570, Mexico}}

\date{2nd August 2010}

\maketitle

\begin{abstract}
We analyse the anomalous properties of specific electronic states in the
Kronig-Penney model with weak compositional and structural disorder.
Using the Hamiltonian map approach, we show that the localisation length
of the electronic states exhibits a resonant effect close to the band
centre and anomalous scaling at the band edges. These anomalies
are akin to the corresponding ones found in the Anderson model with
diagonal disorder.
We also discuss how specific cross-correlations between compositional
and structural disorder can generate an anomalously localised state near
the middle of the energy band.
The tails of this state decay with the same stretched-exponential law
which characterises the band-centre state in the Anderson model with purely
off-diagonal disorder.
\end{abstract}

Pacs: 73.20.Jc, 73.20.Fz, 71.23.An

\section{Introduction}

The interest in one-dimensional random models has been constantly increasing
since the discovery that, contrary to previous beliefs, systems of this class
can undergo a sort of localisation-delocalisation transition when disorder
exhibits specific long-range correlations~\cite{Mou98}.
Nowadays, one-dimensional models with correlated disorder are used in
many fields of physics, including semiconductor superlattices~\cite{Bel99},
bilayered media~\cite{Izr09}, Bose-Einstein condensates~\cite{San07},
transmission in waveguides~\cite{Kuh00} and structures with corrugated
surfaces~\cite{Izr03}, and DNA modelling~\cite{Kro09}.

In many of these problems, the system under study is represented in terms
of an aperiodic Kronig-Penney model.
The original Kronig-Penney model was introduced in the early 1930s to
analyse the band structure of crystalline materials~\cite{Kro31} and
found new applications in the 1980s when it was used to study electronic
states in semiconductor superlattices~\cite{Dav98}.
More recently, the interest in random one-dimensional systems has
spurred investigations of the aperiodic variants of the Kronig-Penney model,
which have been analysed both theoretically and
experimentally~\cite{Bel99,Kuh00,Izr01}.

Because of the simplicity and versatility of the aperiodic Kronig-Penney
model, it is important to obtain as much analytical information as possible
on the structure of its electronic states.
A detailed study of the Kronig-Penney model with weak compositional
and structural disorder was done in~\cite{Her08}, where an analytical
expression for the localisation length of the electronic states was derived.
The formula works very well for most values of the energy, but fails in a
neighbourhood of the band centre and of the band edges, where localisation
anomalies appear.
The first goal of the present work is to understand these anomalies:
using the Hamiltonian map approach~\cite{Izr98}, we show that they have the
same form of the corresponding anomalies which exist in the standard Anderson
model with diagonal disorder at the band centre~\cite{Kap81,Lam84} and at
the band edge~\cite{Der84}.
These anomalies have recently attracted attention because of the related
violations of the single-parameter scaling hypothesis~\cite{Abr79} at
the edge~\cite{Dey00} and at the centre~\cite{Sch03} of the energy band.
It is therefore of some interest to see whether similar anomalies exist
in the random Kronig-Penney model.

After explaining the nature of the anomalies of the localisation length
at the band centre and band edges, we focus our attention on a particular
variant of the random Kronig-Penney model, characterised by specific
cross-correlations between the compositional disorder and the structural
one.
We show that such a Kronig-Penney model corresponds to a Anderson model with
energy-dependent diagonal and off-diagonal disorder. For a specific value
of the energy, the diagonal disorder vanishes and the Kronig-Penney model
becomes equivalent to the Anderson model with purely off-diagonal disorder.
As is known~\cite{Fle77}, at the band centre the Anderson model with purely
diagonal disorder has anomalously localised states, whose existence is
another violation of the single-parameter scaling hypothesis~\cite{Che06}.
Because of the identity of the mathematical equations, similar states
with stretched-exponential tails also exist in the considered variant of
the Kronig-Penney model.

This paper is organised as follows. In Sec.~\ref{standard} we define
the model under study and we summarise the main results concerning the
localisation of the electronic states in the standard case.
In Sec.~\ref{anomal} we discuss the anomalies of the localisation
length emerging for values of the energy close to the band centre
and in the neighbourhood of the band edge.
In Sec.~\ref{stretchexp} we analyse the anomalously localised states
created by particular cross-correlations of the disorder.
The conclusions are drawn in Sec.~\ref{conclu}.

\section{Localisation of the electronic states: the standard case}
\label{standard}

In this section we define the model under study and we briefly derive
the expression for the localisation length valid in the general case.

\subsection{Definition of the model}

We consider a Kronig-Penney model with weak compositional and structural
disorder. The model is defined by the Schr\"{o}dinger equation
\begin{equation}
-\psi^{\prime \prime}(x) +
\sum_{n=-\infty}^{\infty} (U + u_{n}) \delta(x - an - a_{n}) \psi(x) =
q^{2} \psi(x)
\label{kpmodel}
\end{equation}
which describes the motion of an electron of energy $q^{2}$ in a
potential formed by a succession of aperiodically positioned
delta-barriers of random strengths.
To simplify the form of the equations, here and in what follows we use
energy units such that $\hbar^{2}/2m = 1$.
We introduce compositional disorder in model~(\ref{kpmodel}) by assuming
that the barrier strengths display random fluctuations $u_{n}$ around their
mean value $U$.
The structural disorder, on the other hand, is present because the
positions of the barriers are displaced by a random amount $a_{n}$ with
respect to the lattice sites $na$.

For weak disorder, it is enough to specify the statistical properties
of the model in terms of the first two moments of the strength fluctuations
$u_{n}$ and of the relative displacements $\Delta_{n} = a_{n+1} - a_{n}$
(which are more relevant than the absolute displacements $a_{n}$ themselves
for the description of the electronic states).
We assume that both variables have zero average, $\langle u_{n} \rangle =0$
and $\langle \Delta_{n} \rangle = 0$, and that their variances satisfy the
conditions of weak compositional disorder
\begin{displaymath}
\langle u_{n}^{2} \rangle \ll U^{2} ,
\end{displaymath}
and of weak structural disorder
\begin{equation}
\begin{array}{ccc}
\langle \Delta_{n}^{2} \rangle q^{2} \ll 1, &  &
\langle \Delta_{n}^{2} \rangle U^{2} \ll 1 .
\end{array}
\label{weaksd}
\end{equation}
We also consider the normalised binary correlators
\begin{equation}
\begin{array}{ccl}
\chi_{1} (l) & = & \displaystyle
\frac{\langle u_{n} u_{n+l} \rangle}{\langle u_{n}^{2} \rangle} \\
\chi_{2} (l) & = & \displaystyle
\frac{\langle \Delta_{n} \Delta_{n+l} \rangle}
{\langle \Delta_{n}^{2} \rangle} \\
\chi_{3} (l) & = & \displaystyle
\frac{\langle u_{n} \Delta_{n+l} \rangle} {\langle u_{n} \Delta_{n} \rangle} \\
\end{array}
\label{bincor}
\end{equation}
as given functions. We do not attribute any specific form to the
correlators~(\ref{bincor}); we only assume that, because of the spatial
homogeneity in the mean of the model, they are (decreasing) functions
of the difference $l$ of the site indices.
For the sake of simplicity, we restrict our attention to the case in
which the self-correlators $\chi_{1}(l)$ and $\chi_{2}(l)$ are even functions
of $l$, but we do not make the same assumption about the cross-correlator
$\chi_{3}(l)$ (in other words, we suppose that the model is only partially
isotropic).

\subsection{The Hamiltonian map approach}

The Schr\"{o}dinger equation~(\ref{kpmodel}) has the same
form of the dynamical equation of a stochastic oscillator with
frequency $q$ perturbed by a random succession of delta-shaped impulses
(``kicks''). The Hamiltonian of such an oscillator is
\begin{equation}
H = \frac{p^{2}}{2} + \frac{1}{2} \left[ q^{2} - \sum_{n=-\infty}^{\infty}
\left( U + u_{n} \right) \delta \left( t - an - a_{n} \right) \right] x^{2} .
\label{kickosc}
\end{equation}
The mathematical identity of the dynamical equation of the kicked
oscillator~(\ref{kickosc}) with Eq.~(\ref{kpmodel}) allows one to
analyse the properties of the electronic states of the Kronig-Penney
model~(\ref{kpmodel}) in terms of the trajectories of the kicked
oscillator~(\ref{kickosc}): this is the key idea of the Hamiltonian map
approach~\cite{Izr98,Izr95}.
After integrating the Hamiltonian equations of the stochastic oscillator
over the interval between two successive kicks, one obtains the Hamiltonian
map
\begin{equation}
\left( \begin{array}{c}
x_{n+1} \\
p_{n+1} \\
\end{array} \right) =
{\bf T}_{n}
\left( \begin{array}{c}
x_{n} \\
p_{n} \\
\end{array} \right)
\label{hammap1}
\end{equation}
with the transfer matrix
\begin{equation}
{\bf T}_{n} = \left(
\begin{array}{cc}
\displaystyle
\cos \left[q \left( a + \Delta_{n} \right)\right] +
\left( U + u_{n} \right) \frac{1}{q}
\sin \left[ q \left( a + \Delta_{n} \right)\right] &
\displaystyle
\frac{1}{q} \sin \left[q \left( a + \Delta_{n} \right) \right]\\
\displaystyle
-q \sin \left[q \left( a + \Delta_{n} \right) \right]
+\left( U + u_{n} \right)
\cos \left[ q \left( a + \Delta_{n} \right) \right]&
\displaystyle
\cos \left[ q \left( a + \Delta_{n} \right) \right]\\
\end{array}
\right) .
\label{transfer}
\end{equation}
To analyse the trajectories of the Hamiltonian map~(\ref{hammap1}), we
follow the approach proposed in~\cite{Izr01} (see~\cite{Her08} for
details) and we perform the canonical transformation
$(x_{n}, p_{n}) \rightarrow (X_{n}, P_{n})$ of the form
\begin{equation}
\left( \begin{array}{c}
x_{n} \\
p_{n} \\
\end{array} \right) =
\left( \begin{array}{cc}
\displaystyle
\alpha \cos \frac{q a}{2} &
\displaystyle
\frac{1}{q \alpha} \sin
\frac{q a}{2} \\
\displaystyle
- q \alpha \sin \frac{q a}{2} &
\displaystyle
\frac{1}{\alpha} \cos \frac{q a}{2} \\
\end{array} \right)
\left( \begin{array}{c}
X_{n} \\
P_{n} \\
\end{array} \right) .
\label{canonic}
\end{equation}
The parameter $\alpha$ in Eq.~(\ref{canonic}) is defined by the identity
\begin{displaymath}
\alpha^{4} = \frac{1}{q^{2}}
\frac{\displaystyle \sin \left(q a \right)
- \frac{U}{2q} \left[ \cos \left( q a \right) - 1 \right]}
{\displaystyle
\sin \left(q a \right) - \frac{U}{2q}
\left[ \cos \left( q a \right) + 1 \right]} .
\end{displaymath}
The utility of the transformation~(\ref{canonic}) lies in the fact that
the Hamiltonian map~(\ref{hammap1}), when expressed in terms of the new
variables $(X_{n}, P_{n})$, is reduced to a simple phase-space rotation in
the absence of disorder.
The rotation angle $ka$ of the unperturbed map is defined by the identity
\begin{equation}
\cos \left( k a \right) = \cos \left( q a \right) +
\frac{U}{2q} \sin \left( q a \right).
\label{kpbandstructure}
\end{equation}
Eq.~(\ref{kpbandstructure}) determines the band structure of the
Kronig-Penney model~(\ref{kpmodel}) and reveals that the parameter $k$ is
the Bloch wavenumber.

To simplify the form of the Hamiltonian map~(\ref{hammap1}) further, we
switch from the Cartesian coordinates $(X_{n},P_{n})$ to action-angle
variables $(J_{n}, \theta_{n})$ via the standard relations
\begin{displaymath}
\begin{array}{ccl}
X_{n} & = & \sqrt{2 J_{n}} \sin \theta_{n} \\
P_{n} & = & \sqrt{2 J_{n}} \cos \theta_{n} \\
\end{array} .
\end{displaymath}
Within the second-order approximation, the Hamiltonian map~(\ref{hammap1})
can be written in terms of the action-angle variables in the form~\cite{Her08}
\begin{equation}
\begin{array}{ccl}
J_{n+1} & = & D_{n}^{2} J_{n} \\
\theta_{n+1} & = & \displaystyle
\theta_{n} + ka - \frac{1}{2}
\left[1 - \cos \left( 2 \theta_{n} + ka \right) \right] \tilde{u}_{n} 
+ \frac{1}{2} \left[ \zeta - \cos \left( 2 \theta_{n} + 2 ka \right) \right]
\tilde{\Delta}_{n} \\
& + & \displaystyle
\frac{1}{8}
\left[ 2 \sin \left( 2 \theta_{n} + ka \right) - \sin \left( 4 \theta_{n}
+ 2 ka \right) \right] \tilde{u}_{n}^{2} \\
& + & \displaystyle
\frac{1}{8} \left[ 2 \zeta \sin \left( 2 \theta_{n} + 2 ka \right) -
\sin \left( 4 \theta_{n} + 4 ka \right) \right] \tilde{\Delta}_{n}^{2} \\
& + & \displaystyle
\frac{1}{4} \left[ \sin \left( ka \right) -
2 \sin \left( 2 \theta_{n} + 2 ka \right) + \sin \left( 4 \theta_{n} +
3 ka \right) \right] \tilde{u}_{n} \tilde{\Delta}_{n} 
\end{array}
\label{hammap5}
\end{equation}
with the ratio of the action variables being equal to
\begin{equation}
\begin{array}{ccl}
D_{n}^{2} & = & \displaystyle
1 + \sin \left( 2 \theta_{n} + ka \right) \tilde{u}_{n} -
\sin \left( 2 \theta_{n} + 2 ka \right) \tilde{\Delta}_{n} \\
& + & \displaystyle
\frac{1}{2} \left[ 1 - \cos \left( 2 \theta_{n} + k a \right) \right]
\tilde{u}_{n}^{2} +
\frac{1}{2} \left[ 1 - \zeta \cos \left( 2 \theta_{n} + 2 k a \right)
\right] \tilde{\Delta}_{n}^{2} \\
& + & \displaystyle
\left[ \cos \left( 2 \theta_{n} + 2 k a \right) -\cos \left( k a \right)
\right] \tilde{u}_{n} \tilde{\Delta}_{n} .
\end{array}
\label{dnsquared}
\end{equation}
In Eqs.~(\ref{hammap5}) and~(\ref{dnsquared}) we have introduced the
rescaled disorder variables
\begin{eqnarray}
\tilde{u}_{n} =
\frac{\sin \left( qa \right)}{q \sin \left( ka \right)} u_{n} &
\mbox{ and } &
\tilde{\Delta}_{n} = \frac{U}{\sin \left( ka \right)} \Delta_{n} 
\label{resrv}
\end{eqnarray}
and the parameter
\begin{equation}
\zeta = \frac{q \sin \left( ka \right)}{U} \left[ q \alpha^{2} +
\frac{1}{q \alpha^{2}} \right] .
\label{zeta}
\end{equation}

\subsection{The localisation length}

The inverse localisation length for the Kronig-Penney model~(\ref{kpmodel})
is defined as
\begin{displaymath}
l^{-1}_{\rm loc} = \lim_{N \rightarrow \infty} \frac{1}{Na} \sum_{n=1}^{N}
\log \left|\frac{\psi_{n+1}}{\psi_{n}}\right| .
\end{displaymath}
In the dynamical picture, $l^{-1}_{\rm loc}$ is equivalent to the Lyapunov
exponent of the Hamiltonian map~(\ref{hammap1}),
\begin{equation}
\lambda = \lim_{N \rightarrow \infty} \frac{1}{Na} \sum_{n=1}^{N}
\log \left| \frac{x_{n+1}}{x_{n}} \right| =
\lim_{N \rightarrow \infty} \frac{1}{Na} \sum_{n=1}^{N}
\log \left( D_{n} \right) + \lim_{N \rightarrow \infty} \frac{1}{Na}
\sum_{n=1}^{N} \log \left( R_{n} \right),
\label{twotermlyap}
\end{equation}
with $D_{n}$ being defined by Eq.~(\ref{dnsquared}) and
\begin{equation}
R_{n} =  \left| \frac{\sqrt{\zeta + 1} \left[ 1 + \cos \left(qa \right) \right]
\sin \theta_{n+1} + \sqrt{\zeta - 1} \sin \left( qa \right) \cos \theta_{n+1}}
{\sqrt{\zeta + 1} \left[ 1 + \cos \left(qa \right) \right]
\sin \theta_{n} + \sqrt{\zeta - 1} \sin \left( qa \right) \cos \theta_{n}}
\right| .
\label{thetaratio}
\end{equation}
As discussed in Sec.~\ref{anomal}, in the weak-disorder case the invariant
distribution for the angular variable is either flat (the normal case) or
presents a moderate modulation (as happens at the band centre and
for the other cases in which the Bloch wavenumber is a rational multiple
of $\pi/a$). Only at the band edge the invariant distribution is strongly
modulated even for weak disorder.
Therefore as a function of $n$ the ratio~(\ref{thetaratio}) can be expected
to oscillate around a unitary value, unlike the ratio $D_{n}$ which is
larger than one on average because of the exponential increase of the
action variable. 
As a consequence, away from the band edge, one can drop the second term in
the right-hand side of Eq.~(\ref{twotermlyap}) and compute the inverse
localisation length as
\begin{equation}
\lambda = \lim_{N \rightarrow \infty} \frac{1}{Na} \sum_{n=1}^{N}
\log \left( D_{n} \right) = \langle \log D_{n} \rangle.
\label{lyap1}
\end{equation}
Within the second-order approximation one can expand the logarithm
in Eq.~(\ref{lyap1}) and obtain
\begin{equation}
\begin{array}{ccl}
\lambda & = & \displaystyle
\frac{1}{2a} \biggl\langle
\sin \left( 2 \theta_{n} + ka \right) \tilde{u}_{n} -
\sin \left( 2 \theta_{n} + 2 ka \right) \tilde{\Delta}_{n} \\
& + & \displaystyle
\frac{1}{4} \left[ 1 - 2 \cos \left( 2 \theta_{n} + ka \right) +
\cos \left( 4 \theta_{n} + 2 ka \right) \right] \tilde{u}_{n}^{2} \\
& + & \displaystyle
\frac{1}{4} \left[ 1 - 2 \zeta \cos \left( 2 \theta_{n} + 2 ka \right)
+ \cos \left( 4 \theta_{n} +  4ka \right) \right] \tilde{\Delta}_{n}^{2} \\
& - & \displaystyle
\frac{1}{2} \left[ \cos \left( ka \right) - 2 \cos \left( 2 \theta_{n} +
2 ka \right) + \cos \left( 4 \theta_{n} + 3 ka \right) \right]
\tilde{u}_{n} \tilde{\Delta}_{n}  \biggr\rangle \\
\end{array}
\label{logd}
\end{equation}
In the general case, the average over the angle variable in the right-hand
side of Eq.~(\ref{logd}) can be computed using a flat distribution
$\rho (\theta) = 1/(2 \pi)$.
In fact, as the Hamiltonian map~(\ref{hammap5}) shows, the angle variable
has a fast dynamics compared to the action variable and quickly assumes a
uniform distribution in the interval $[0,2 \pi]$.
In this way, following the steps of Ref.~\cite{Her08}, one obtains that the
inverse localisation length~(\ref{lyap1}) has the form
\begin{equation}
\lambda = \frac{1}{8a} \left[ \langle \tilde{u}_{n}^{2} \rangle W_{1}(ka)
+ \langle \tilde{\Delta}_{n}^{2} \rangle W_{2}(ka)
- 2 \langle \tilde{u}_{n} \tilde{\Delta}_{n} \rangle W_{3}(ka) \right]
\label{invloc0}
\end{equation}
with
\begin{displaymath}
\begin{array}{ccl}
W_{1}(ka) & = & \displaystyle
\sum_{l=-\infty}^{+\infty} \chi_{1}(l) \cos \left( 2kal \right), \\
W_{2}(ka) & = & \displaystyle
\sum_{l=-\infty}^{+\infty} \chi_{2}(l) \cos \left( 2kal \right), \\
W_{3}(ka) & = & \displaystyle
\sum_{l=-\infty}^{+\infty} \chi_{3}(l) \cos \left[ ka \left( 2l+1 \right)
\right]. \\
\end{array}
\end{displaymath}
Eq.~(\ref{invloc0}) constitutes the standard expression of the inverse
localisation length of the electronic states in the Kronig-Penney
model~(\ref{kpmodel}).
Note that expression~(\ref{invloc0}) is slightly more general than the
corresponding formula given in~\cite{Her08} because we have dropped the
assumption that the cross-correlator $\chi_{3}(k)$ is an even function of
$k$. The difference between the two expressions shows up in the last term
of Eq.~(\ref{invloc0}) and in the definition of the $W_{3}(ka)$ power
spectrum. When $\chi_{3}(k)= \chi_{3}(-k)$, Eq.~(\ref{invloc0}) reduces
to the inverse localisation length derived in~\cite{Her08}.

The inverse localisation length~(\ref{invloc0}) works well for almost
all values of the energy, but fails close to the band centre and at the
band edges, where anomalous effects appear. The nature of these anomalies
is discussed in the next section.

\section{The band-centre and band-edge anomalies of the localisation length}
\label{anomal}

As mentioned in the previous section, the derivation of
formula~(\ref{invloc0}) rests on the crucial assumption that the angle
variable of the map~(\ref{hammap5}) has a {\em uniform} invariant
distribution.
This assumption is generally justified on the grounds that the angle
variable evolves much faster than the action variable and quickly sweeps
the whole interval $[0,2\pi]$.
One should notice, however, that this argument cannot be applied when the
Bloch wavevector is a rational multiple of $\pi/a$, because in this case
the noiseless angular map has periodic orbits, whose effect persists in
the form of a modulation of the invariant measure when a weak noise is
switched on .
For weak disorder, and within the limits of the second-order approximation,
this implies that the general formula~(\ref{invloc0}) cannot be applied
when the Bloch vector takes values $k a \simeq 0$ or $ka \simeq \pm \pi$,
i.e., at the edges of the energy band, and for $k = \pm \pi/2a$, i.e.,
for the energy $[q(\pi/2a)]^{2}$ which lies close to the band centre.
In principle, one should expect anomalies for all Bloch vectors of the
form $k a = \pi/n$ with $|n| > 2$. In practice, however, in these cases
the invariant distribution of the angular variable is modified by a
perturbative term proportional to $\cos (2n \theta)$ or $\sin (2 n \theta)$
and this does not affect the outcome of the average in Eq.~(\ref{logd}),
which contains only second- and fourth-order harmonics of $\theta$
(the same conclusion applies to the Anderson model, as discussed in
Ref.~\cite{Lam84}).

In conclusion, within the second-order approximation anomalies are found
only at the band edge ($ka \simeq 0$ or $ka \simeq \pm \pi$) or
close to the middle of the band ($ka = \pi/2$).
For these special values of $k$ the assumption of a flat invariant
distribution must be abandoned and the specific form of $\rho(\theta)$ has
to be determined before one can compute the localisation length of the
electronic states.
The non-uniform distribution of the angular variable produces deviations
from the standard formula~(\ref{invloc0}) of the inverse localisation
length; this section is devoted to the analysis of these anomalies.

For the sake of simplicity, in this section we will restrict our attention
to the case of {\em uncorrelated} disorder.
In other words, we will consider two successions of random variables
$\{ u_{n} \}$ and $\{ \Delta_{n} \}$ such that
\begin{displaymath}
\begin{array}{ccc}
\langle u_{n} u_{k} \rangle = \langle u_{n}^{2} \rangle \delta_{n k}, &
\langle \Delta_{n} \Delta_{k} \rangle = 
\langle \Delta_{n}^{2} \rangle \delta_{n k}, &
\langle u_{n} \Delta_{k} \rangle = 0 .
\end{array}
\end{displaymath}
In this case the standard expression~(\ref{invloc0}) of the inverse
localisation length reduces to the simple form
\begin{equation}
\lambda = \frac{1}{8a} \left[ \langle \tilde{u}_{n}^{2} \rangle +
\langle \tilde{\Delta}_{n}^{2} \rangle \right] =
\frac{1}{8a \sin^{2}(ka)} \left[ \frac{\sin^{2}(qa)}{q^{2}}
\langle u_{n}^{2} \rangle + U^{2} \langle \Delta_{n}^{2} \rangle \right] .
\label{uncorlyap}
\end{equation}

\subsection{The anomaly near the middle of the energy band}

We will first consider the anomaly for $ka= \pm \pi/2$. The corresponding
energy lies close to middle of the band and therefore we will often speak,
somewhat loosely, of band-centre anomaly.
By assigning the value $k = \pi/2a$ to the Bloch vector and by taking into
account that the noise-angle correlators vanish for uncorrelated disorder,
one can reduce the general expression~(\ref{logd}) to the simpler form
\begin{equation}
\begin{array}{ccl}
\lambda & = & \displaystyle
\frac{1}{8a} \left\{ \left[ 1 + 2 \langle \sin \left( 2 \theta_{n} \right) 
\rangle - \langle \cos \left( 4 \theta_{n} \right) \rangle \right]
\langle \tilde{u}_{n}^{2} \rangle \right. \\
& + & \displaystyle \left.
\left[ 1 + 2 \zeta \langle \cos \left( 2 \theta_{n} \right) \rangle
+ \langle \cos \left( 4 \theta_{n} \right) \rangle \right]
\langle \tilde{\Delta}_{n}^{2} \rangle \right\} .
\end{array}
\label{lyap}
\end{equation}

\begin{figure}[thb]
\begin{center}
\epsfig{file=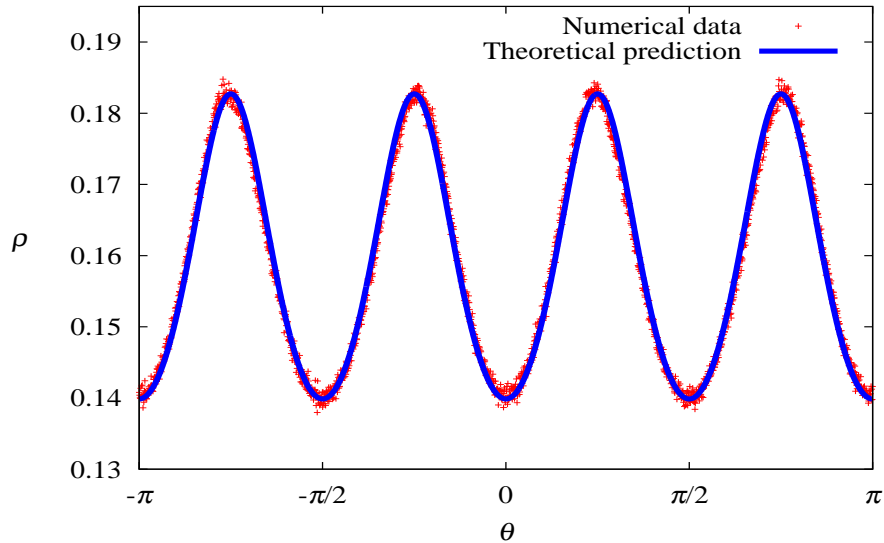,width=5in,height=3in}
\caption{(Colour on line)
Invariant distribution $\rho$ versus $\theta$. The solid line
corresponds to the prediction of Eq.~(\ref{invmea}), while points
represent numerical results. \label{rho}}
\end{center}
\end{figure}

To determine the form of the invariant distribution $\rho(\theta)$,
we follow the method  introduced in~\cite{Izr98} and we consider the map
for the angle variable in Eq.~(\ref{hammap5}) which, for uncorrelated
disorder and $k = \pi/2a$, simplifies to
\begin{displaymath}
\begin{array}{ccl}
\theta_{n+1} & = & \displaystyle
\theta_{n} + \frac{\pi}{2} -
\frac{1}{2} \left[ 1 + \sin \left( 2 \theta_{n} \right) \right]
\tilde{u}_{n} +
\frac{1}{2} \left[ \zeta + \cos \left( 2 \theta_{n} \right) \right]
\langle \tilde{\Delta}_{n} \rangle \\
& + & \displaystyle
\frac{1}{8} \left[ 2 \cos \left( 2 \theta_{n} \right) +
\sin \left( 4 \theta_{n} \right) \right]
\langle \tilde{u}_{n}^{2} \rangle -
\frac{1}{8} \left[ 2 \zeta \sin \left( 2 \theta_{n} \right) +
\sin \left( 4 \theta_{n} \right) \right]
\langle \tilde{\Delta}_{n}^{2} \rangle .
\end{array}
\end{displaymath}
To get rid of the constant drift term, one can consider the fourth
iterate of this map, i.e.,
\begin{displaymath}
\begin{array}{ccl}
\theta_{n+4} & = & \displaystyle
\theta_{n} + \frac{1}{2} \sin \left( 4 \theta_{n} \right)
\left( \langle \tilde{u}_{n}^{2} \rangle -
\langle \tilde{\Delta}_{n}^{2} \rangle \right) \\
& - &\displaystyle
\frac{1}{2} \left[ 1 + \sin \left( 2 \theta_{n} \right) \right]
\left( \tilde{u}_{n} + \tilde{u}_{n+2} \right)
-\frac{1}{2} \left[ 1 - \sin \left( 2 \theta_{n} \right) \right]
\left( \tilde{u}_{n+1} + \tilde{u}_{n+3} \right) \\
& + & \displaystyle
\frac{1}{2} \left[ \zeta + \cos \left( 2 \theta_{n} \right) \right]
\left( \tilde{\Delta}_{n} + \tilde{\Delta}_{n+2} \right) +
\frac{1}{2} \left[ \zeta - \cos \left( 2 \theta_{n} \right) \right]
\left( \tilde{\Delta}_{n+1} + \tilde{\Delta}_{n+3} \right). \\
\end{array}
\end{displaymath}
Going to the continuum limit, one can replace this map with the It\^{o}
stochastic differential equation
\begin{equation}
\begin{array}{ccl}
{\mathrm d} \theta & = & \displaystyle
\frac{1}{2} \sin \left( 4 \theta \right)
\left( \langle \tilde{u}_{n}^{2} \rangle -
\langle \tilde{\Delta}_{n}^{2} \rangle \right) {\mathrm d}t \\
& - & \displaystyle
\sqrt{\frac{\langle \tilde{u}_{n}^{2} \rangle}{2}}
\left[ 1 + \sin \left( 2 \theta \right) \right] {\mathrm d}W_{1}
- \sqrt{\frac{\langle \tilde{u}_{n}^{2} \rangle}{2}}
\left[ 1 - \sin \left( 2 \theta \right) \right] {\mathrm d}W_{2} \\
& + & \displaystyle
\sqrt{\frac{\langle \tilde{\Delta}_{n}^{2} \rangle}{2}}
\left[ \zeta + \cos \left( 2 \theta \right) \right] {\mathrm d}W_{3}
+ \sqrt{\frac{\langle \tilde{\Delta}_{n}^{2} \rangle}{2}}
\left[ \zeta - \cos \left( 2 \theta \right) \right] {\mathrm d}W_{4} \\
\end{array}
\label{ito}
\end{equation}
where $W_{1}(t), \ldots, W_{4}(t)$ represent four independent Wiener
processes.
Given an initial condition $\theta(t_{0}) = \theta_{0}$, the It\^{o}
equation~(\ref{ito}) defines a stochastic process $\theta(t)$, whose
conditional probability $p(\theta, t| \theta_{0}, t_{0}) = p$ can
be obtained by solving the associated Fokker-Planck equation
\begin{equation}
\begin{array}{ccl}
\displaystyle
\frac{\partial p}{\partial t} & = & \displaystyle
\frac{1}{2} \left( \langle \tilde{\Delta}_{n}^{2} \rangle -
\langle \tilde{u}_{n}^{2} \rangle \right) \frac{\partial}{\partial \theta}
\left[ \sin \left( 4 \theta \right) p \right] \\
& + & \displaystyle
\frac{1}{4} \frac{\partial^{2}}{\partial \theta^{2}}
\left\{ \left[ \left( 3 \langle \tilde{u}_{n}^{2} \rangle +
\left( 2 \zeta^{2} + 1 \right) \langle \tilde{\Delta}_{n}^{2} \rangle \right)
+ \left( \langle \tilde{\Delta}_{n}^{2} \rangle -
\langle \tilde{u}_{n}^{2} \rangle \right) \cos \left( 4 \theta \right)
\right] p \right\} \\
\end{array}
\label{fp}
\end{equation}
with the initial condition $p(\theta, t_{0}| \theta_{0}, t_{0}) =
\delta (\theta - \theta_{0})$~\cite{Gar04}.
The stationary solution of Eq.~(\ref{fp}) which satisfies the conditions
of normalisation and periodicity is~\cite{Izr98}
\begin{equation}
\rho(\theta) = \frac{\sqrt{A + |B|}}{4 {\bf K} \left( C \right)}
\frac{1}{\sqrt{A - B \cos \left( 4 \theta \right)}} .
\label{invmea}
\end{equation}
In Eq.~(\ref{invmea}) ${\bf K}(C)$ is the complete elliptic integral of the
first kind and we have introduced the constants
\begin{eqnarray*}
A = 3 \langle \tilde{u}_{n}^{2} \rangle + \left( 2 \zeta^{2} + 1 \right)
\langle \tilde{\Delta}_{n}^{2} \rangle, &
\mbox{ \hspace{0.5cm} } &
B = \langle \tilde{u}_{n}^{2} \rangle - \langle \tilde{\Delta}_{n}^{2} \rangle
\end{eqnarray*}
and
\begin{equation}
C = \sqrt{\frac{2 |B|}{A + |B|}} = \sqrt{ \frac{ 2 \left|
\langle \tilde{u}_{n}^{2} \rangle - \langle \tilde{\Delta}_{n}^{2} \rangle
\right|}{\left|
\langle \tilde{u}_{n}^{2} \rangle - \langle \tilde{\Delta}_{n}^{2} \rangle
\right| + 3 \langle \tilde{u}_{n}^{2} \rangle +
\left( 2 \zeta^{2} + 1 \right) \langle \tilde{\Delta}_{n}^{2} \rangle}} .
\label{c}
\end{equation}
Eq.~(\ref{invmea}) shows that, as expected, when the Bloch vector
takes the value $k = \pi/2a$, the invariant measure has period $\pi/2$.
The numerical computations agree well with formula~(\ref{invmea}) as can
be seen in Fig.~\ref{rho}.
The data represented in Fig.~\ref{rho} were obtained for mean field
$U = 8$ and disorder strengths $\sqrt{\langle u_{n}^{2} \rangle} =
\sqrt{\langle \Delta_{n}^{2} \rangle} = 0.02$. Here, and in the rest of
the paper, we present numerical data which were obtained for energy values
within the first band.

\begin{figure}[thb]
\begin{center}
\epsfig{file=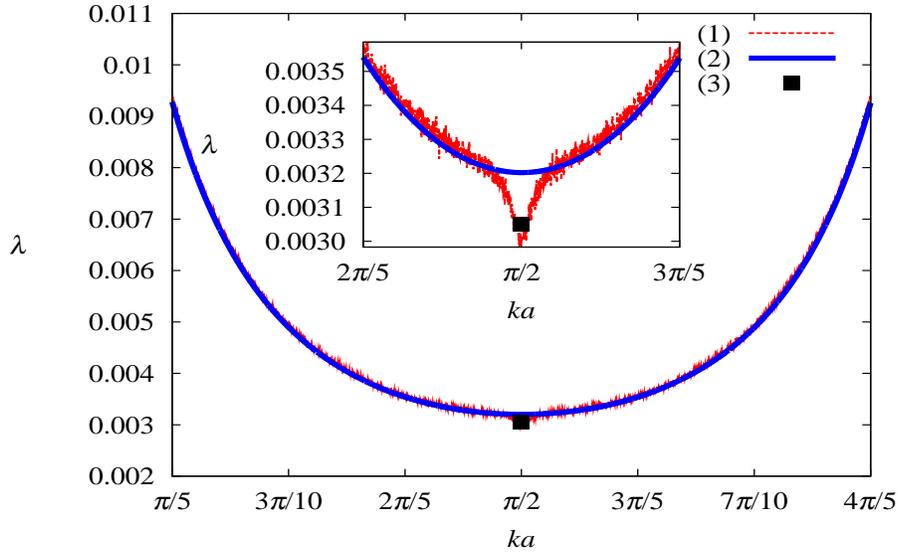,width=5in,height=3in}
\caption{(Colour on line)
Inverse localisation length $\lambda$ versus $ka$ (uncorrelated disorder).
The dashed line (1) represents numerically obtained values; the solid line
(2) corresponds to formula~(\ref{uncorlyap}); the symbol (3) represents the
anomalous value~(\ref{anolyap}).
The inset shows a close-up of the anomaly.
\label{bc_anomal}}
\end{center}
\end{figure}

The knowledge of the invariant distribution~(\ref{invmea}) makes possible
to compute the Lyapunov exponent~(\ref{lyap}).
The averages of the functions of argument $2 \theta$ vanish, but the the
average of $\cos (4\theta)$ does not and gives rise to the anomaly of the
localisation length.
After some algebra, one obtains that the inverse localisation length
for $ka = \pi/2$ is
\begin{equation}
\begin{array}{ccl}
\lambda & = & \displaystyle
\frac{1}{8a} \bigg\{ \left[ \left| \langle \tilde{u}_{n}^{2} \rangle -
\langle \tilde{\Delta}_{n}^{2} \rangle \right| + 3
\langle \tilde{u}_{n}^{2} \rangle + \left( 2 \zeta^{2} + 1 \right)
\langle \tilde{\Delta}_{n}^{2} \rangle \right]
\frac{{\bf E}(C)}{{\bf K}(C)} \\
& - & \displaystyle
2 \left( \langle \tilde{u}_{n}^{2} \rangle + \zeta^{2}
\langle \tilde{\Delta}_{n}^{2} \rangle \right) \bigg\} \\
\end{array}
\label{anolyap}
\end{equation}
where ${\bf E}(C)$ is the complete elliptic integral of the second kind and
the argument $C$ is defined by Eq.~(\ref{c}).
The numerical computations confirm the existence of an anomaly for $ka = \pi/2$
as can be seen from the data represented in Fig.~\ref{bc_anomal} which
show a small but clear deviation from the value of the localisation
length predicted by the standard formula~(\ref{uncorlyap}).
The numerically computed inverse localisation length for $ka = \pi/2$,
on the other hand, matches well the theoretical value~(\ref{anolyap}).
We observe that the numerical data represented in Fig.~\ref{bc_anomal}
were obtained for a specific realisation of the disorder; when different
disorder realisations are considered, the discrepancy between the numerical
value of the Lyapunov exponent for $ka = \pi/2$ and the predicted
result~(\ref{anolyap}) fluctuates slightly around zero, always assuming
small values as in Fig.~\ref{bc_anomal}.

In conclusion, in the Kronig-Penney model~(\ref{kpmodel}) a resonance
effect occurs for $k=\pi/2a$ and produces an anomaly of the localisation
length. The effect has the same nature of the band-centre anomaly found in
the Anderson model with diagonal disorder~\cite{Kap81,Izr98}.

\begin{figure}[thb]
\begin{center}
\epsfig{file=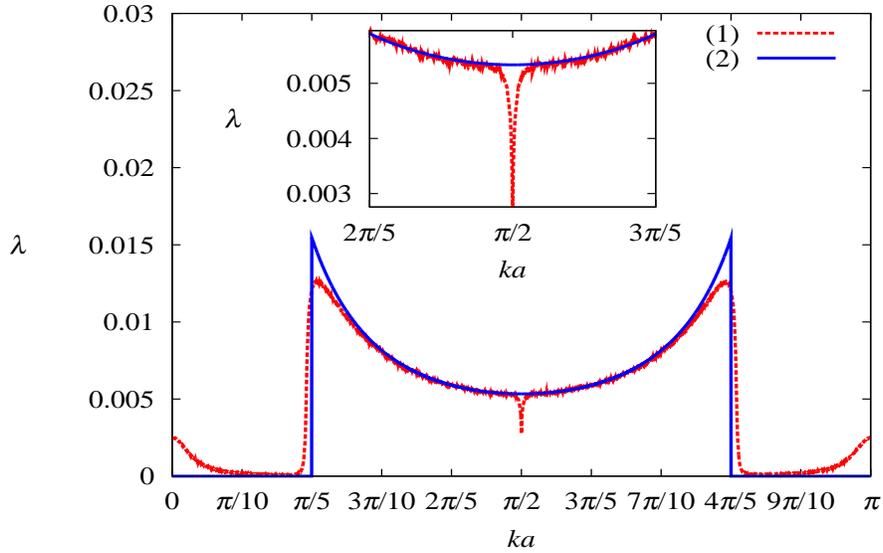,width=5in,height=3in}
\caption{(Colour on line)
Inverse localisation length $\lambda$ versus $ka$ (self-correlated
disorder). The dashed line (1) represents numerically computed data, while
the solid line (2) corresponds  to the values predicted by Eq.~(\ref{invloc0}).
The inset shows the anomaly in detail.
\label{anomalycorr}}
\end{center}
\end{figure}

We remark that, although our analytical results are restricted to
the case of uncorrelated disorder, we found numerical evidence that
disorder correlations can enhance the anomaly of the localisation length
near the middle of the energy band, in agreement with the theoretical
conclusions of Ref.~\cite{Tit05}.
As an example, we can consider the case of structural and compositional
disorder with self-correlations of the form
\begin{equation}
\chi_{1}(l) = \chi_{2}(l) = \left\{
\begin{array}{lcc}
1 & \mbox{ if } & l = 0 \\
\displaystyle
-\frac{5}{3\pi l} \sin \left( \frac{2}{5} \pi l \right) &
\mbox{ if } & |l| > 0 \\
\end{array} \right.
\label{lrcorr}
\end{equation}
and no cross-correlations, $\chi_{3}(l) = 0$.
The long-range correlations of the form~(\ref{lrcorr}) create mobility
edges at $k = \pi/5a$ and $k = 4\pi/5a$ (see~\cite{Her08} for details).
Fig.~\ref{anomalycorr} represents the numerical data obtained for this kind of
disorder with $U=8$ and $\sqrt{\langle u_{n}^{2} \rangle} =
\sqrt{\langle \Delta_{n}^{2} \rangle} = 0.02$.
The data clearly show an enhanced anomaly at $k = \pi/2a$ with respect
to the case of totally uncorrelated disorder. Adding cross-correlations
does not introduce any significant modification to the picture.

We would like to stress that, although the previous example shows how
correlations of the disorder can enhance the anomaly for $k = \pi/2a$,
not all correlations produce the same effect.
This can be appreciated in the case analysed in Sec.~\ref{stretchexp},
in which the resonance effect for $k = \pi/2a$ is shadowed by a different
kind of anomaly generated by specific cross-correlations for a value of $k$
close, but not identical, to $\pi/2a$.

\subsection{The band-edge anomaly}

We now turn our attention to the anomaly for $ka = \varepsilon \to 0^{+}$,
i.e., in the neighbourhood of the band edge. Because of the similarity
with the band-edge anomaly in the Anderson model, we can apply the
method used in~\cite{Izr98} to the present case.
We first derive the form taken by the map~(\ref{hammap5}) at the band
edge.
For $ka = \varepsilon \to 0^{+}$ the rescaled random variables~(\ref{resrv})
can be approximated as
\begin{eqnarray*}
\tilde{u}_{n} = \frac{\sin(q_{0}a)}{q_{0}} \frac{u_{n}}{\varepsilon} +
\ldots & \mbox{ and } &
\tilde{\Delta}_{n} = U \frac{\Delta_{n}}{\varepsilon} + \ldots,
\end{eqnarray*}
where we have introduced the symbol $q_{0} = q(k=0)$.
In the same limit, the parameter~(\ref{zeta}) reduces to
\begin{equation}
\zeta = \sqrt{1 + \frac{4q^{2}}{U^{2}} \sin^{2}(ka)} =
1 + \frac{2q_{0}^{2}}{U^{2}} \varepsilon^{2} + \ldots
\label{bezeta}
\end{equation}
Taking into account these approximations, one can write the Hamiltonian
map~(\ref{hammap5}) in the form
\begin{equation}
\begin{array}{ccl}
J_{n+1} & = & D_{n}^{2} J_{n} \\
\theta_{n+1} & = & \displaystyle
\theta_{n} + \varepsilon - \frac{\xi_{n}}{\varepsilon}
\sin^{2}(\theta_{n}) + \frac{\sigma^{2}}{\varepsilon^{2}} \sin^{3} \left(
\theta_{n} \right) \cos \left( \theta_{n} \right) + \ldots
\end{array}
\label{bemap}
\end{equation}
with
\begin{equation}
D_{n} = 1 - \frac{\xi_{n}}{\varepsilon} \sin \left( \theta_{n} \right)
\cos \left( \theta_{n} \right) + \frac{\sigma^{2}}{2\varepsilon^{2}} 
\sin^{4} \left( \theta_{n} \right) + \ldots
\label{bedn}
\end{equation}
In Eqs.~(\ref{bemap}) and~(\ref{bedn}), the symbol $\xi_{n}$ represents
the linear combination of structural and compositional disorder
\begin{displaymath}
\xi_{n} = U \Delta_{n} - \frac{\sin (q_{0}a)}{q_{0}} u_{n}
\end{displaymath}
with zero average, $\langle \xi_{n} \rangle = 0$, and variance
\begin{displaymath}
\langle \xi_{n}^{2} \rangle = \sigma^{2} =
U^{2} \langle \Delta_{n}^{2} \rangle + \frac{\sin^{2}(q_{0}a)}{q_{0}^{2}}
\langle u_{n}^{2} \rangle .
\end{displaymath}

Going to the continuum limit, one can replace the angular map in
Eq.~(\ref{bemap}) with the stochastic It\^{o} equation
\begin{equation}
{\mathrm d} \theta = \left[ \varepsilon + \frac{\sigma^{2}}{\varepsilon^{2}}
\sin^{3} \left( \theta \right) \cos \left( \theta \right) \right] {\mathrm d}t
+ \frac{\sqrt{\sigma^{2}}}{\varepsilon} \sin^{2} \left( \theta \right)
{\mathrm d}W
\label{beito}
\end{equation}
whose associated Fokker-Planck equation
\begin{equation}
\frac{\partial p}{\partial t} = - \frac{\partial}{\partial \theta}
\left\{ \left[ \varepsilon + \frac{\sigma^{2}}{\varepsilon^{2}}
\sin^{3} (\theta) \cos (\theta) \right] p \right\} +
\frac{1}{2} \frac{\partial^{2}}{\partial \theta^{2}} \left[
\frac{\sigma^{2}}{\varepsilon^{2}} \sin^{4} (\theta) p \right]
\label{befp1}
\end{equation}
gives the conditional probability $p(\theta, t | \theta_{0}, t_{0}) = p$
for the stochastic process $\theta(t)$~\cite{Gar04}.
By introducing the rescaled time
\begin{displaymath}
\tau = \frac{\sigma^{2}}{\varepsilon^{2}} t ,
\end{displaymath}
one can cast the Fokker-Planck equation~(\ref{befp1}) in the form
\begin{equation}
\frac{\partial p}{\partial \tau} = - \frac{\partial}{\partial \theta}
\left\{ \left[ \varkappa + \sin^{3} (\theta) \cos (\theta) \right] p \right\} +
\frac{1}{2} \frac{\partial^{2}}{\partial \theta^{2}} \left[
\frac{\sigma^{2}}{\varepsilon^{2}} \sin^{4} (\theta) p \right]
\label{befp2}
\end{equation}
which contains the noise intensity $\sigma^{2}$ and the distance from the
band edge $\varepsilon$ combined in the {\em single} scaling parameter
\begin{displaymath}
\varkappa = \frac{\varepsilon^{3}}{\sigma^{2}}.
\end{displaymath}
The invariant distribution $\rho(\theta)$ is the stationary solution of
the Fokker-Planck~(\ref{befp2}) which is normalisable and satisfies the
periodicity condition $\rho(\theta + \pi) = \rho(\theta)$.
The solution possessing these features is~\cite{Der84, Izr98}
\begin{equation}
\rho(\theta) = \frac{1}{N(\varkappa)}
\frac{e^{-f(\theta)}}{\sin^{2}(\theta)} \int_{\theta}^{\pi}
\frac{e^{f(\phi)}}{\sin^{2}(\phi)} {\mathrm d}\phi
\label{berho}
\end{equation}
with
\begin{displaymath}
f(\theta) = 2 \varkappa \left[ \frac{1}{3} \cot^{3} (\theta) +
\cot(\theta) \right]
\end{displaymath}
and
\begin{displaymath}
N(\varkappa) = \sqrt{\frac{2 \pi}{\varkappa}} \int_{0}^{\infty}
\frac{1}{\sqrt{x}} \exp \left[ - 2 \varkappa \left( \frac{x^{3}}{12} + x
\right) \right] {\mathrm d}x .
\end{displaymath}
The integral representation~(\ref{berho}) defines the invariant measure in
the interval $[0,\pi]$; $\rho(\theta)$ can be extended outside of this
interval via the periodicity condition $\rho(\theta + \pi) = \rho(\theta)$.

\begin{figure}[thb]
\begin{center}
\epsfig{file=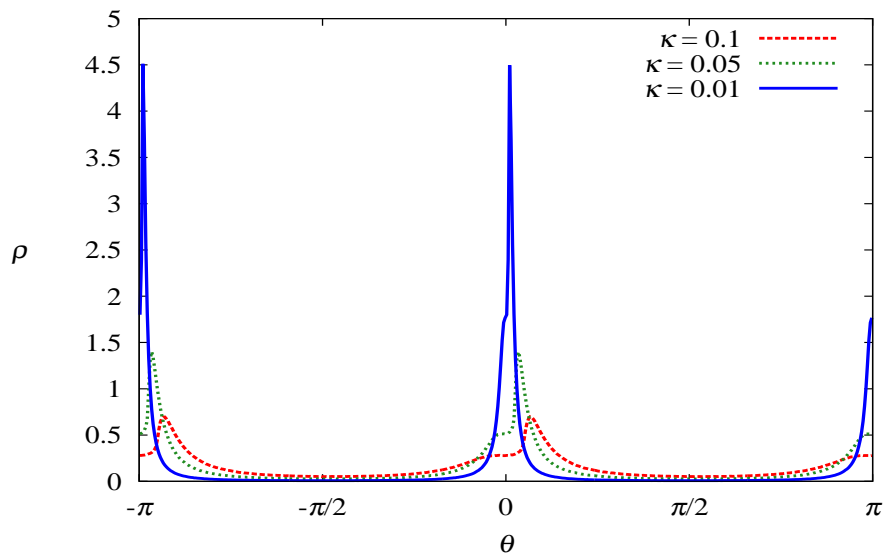,width=5in,height=3in}
\caption{(Colour on line)
Invariant distribution $\rho$ versus $\theta$. The legend shows
the value of $\varkappa$ corresponding to each line.
\label{beinvmea}}
\end{center}
\end{figure}

To obtain a qualitative understanding of the behaviour of the invariant
distribution~(\ref{berho}), it is useful to consider its values at the
edges and at the centre of the $[0,\pi]$ interval.
For $\theta \to 0^{+}$ and $\theta \to \pi^{-}$ one has
\begin{displaymath}
\rho (\theta) \sim
\left\{ \begin{array}{lcl}
\displaystyle
\frac{1}{\varkappa^{5/3}} & \mbox{ if } & \varkappa \to 0 \\
\displaystyle
\frac{1}{2 \pi} & \mbox{ if } & \varkappa \to \infty \\
\end{array} \right. ,
\end{displaymath}
while for $\theta \to \pi/2$ the invariant distribution behaves as
\begin{displaymath}
\rho (\theta) \sim
\left\{ \begin{array}{lcl}
\displaystyle
\varkappa^{1/3} & \mbox{ if } & \varkappa \to 0 \\
\displaystyle
\frac{1}{2 \pi} & \mbox{ if } & \varkappa \to \infty \\
\end{array} \right. .
\end{displaymath}
These equations show that, when the energy moves closer to the band edge on
the scale defined by the disorder strength, i.e., for $\varkappa
\to 0$, the invariant distribution develops two pronounced maxima for
$\theta \sim 0$ and $\theta \sim \pi$.
This conclusion is supported by direct numerical computation of the
invariant distribution, as shown by Fig.~\ref{beinvmea}.
The data in Fig.~\ref{beinvmea} were obtained for mean field $U=8$ and
disorder strengths $\sqrt{\langle u_{n}^{2} \rangle} =
\sqrt{\langle \Delta_{n}^{2} \rangle} = 0.02$.

The assumption of a uniform or slightly modulated distribution must therefore
be radically dropped and this entails that one can no longer neglect the
logarithm of the ratio~(\ref{thetaratio}) in the expression~(\ref{twotermlyap})
of the inverse localisation length.
Taking into account Eq.~(\ref{bezeta}), in the neighbourhood of the band
edge the inverse localisation length~(\ref{twotermlyap}) can be reduced to
the form
\begin{equation}
\lambda \simeq \frac{1}{a}
\Bigg\langle \log \left[ D_{n} \left| \frac{\sin \left( \theta_{n+1} \right)}
{\sin \left( \theta_{n} \right)} \right| \right] \Bigg\rangle.
\label{belambda}
\end{equation}
On the other hand, using the angular map in Eq.~(\ref{bemap}), one
can write the sine ratio in Eq.~(\ref{belambda}) as
\begin{equation}
\begin{array}{ccl}
\displaystyle
\frac{\sin \left( \theta_{n+1} \right)}{\sin \left( \theta_{n} \right)}
& = & \displaystyle
1 + \varepsilon \cot \left( \theta_{n} \right) +
\frac{\xi_{n}}{\varepsilon} \sin \left( \theta_{n} \right)
\cos \left( \theta_{n} \right) \\
& + & \displaystyle
\frac{\sigma^{2}}{\varepsilon^{2}}
\left[ \sin^{2}\left( \theta_{n} \right) \cos^{2}\left( \theta_{n} \right)
- \frac{1}{2} \sin^{4} \left( \theta_{n} \right) \right] + \ldots
\end{array}
\label{besineratio}
\end{equation}
Substituting the approximate identities~(\ref{bedn}) and~(\ref{besineratio})
in Eq.~(\ref{belambda}), one obtains
\begin{displaymath}
\lambda \simeq \frac{\varepsilon}{a} \Big\langle
\cot \left( \theta_{n} \right) \Big\rangle .
\end{displaymath}
The average can now be computed using the invariant distribution~(\ref{berho});
the final result is~\cite{Izr98}
\begin{equation}
\lambda = \frac{\varepsilon}{2a}
\frac{\displaystyle \int_{0}^{\infty} x^{1/2} \exp \left[ -2 \varkappa
\left( \frac{x^{3}}{12} + x \right) \right] {\mathrm d}x}
{\displaystyle \int_{0}^{\infty} x^{-1/2} \exp \left[ -2 \varkappa
\left( \frac{x^{3}}{12} + x \right) \right] {\mathrm d}x} .
\label{beinvloc}
\end{equation}
Away from the band-edge on the length scale set by the disorder
strength, i.e., for $\varkappa \to \infty$, Eq.~(\ref{beinvloc})
reduces to
\begin{displaymath}
\lambda \simeq \frac{\sigma^{2}}{8 a \varepsilon^{2}} ,
\end{displaymath}
which coincides with the form of the standard expression~(\ref{uncorlyap})
in the limit $ka = \varepsilon \to 0^{+}$.
On the other hand, close to the band-edge (on the length scale defined
by the disorder strength), i.e., for $\varkappa \to 0$,
Eq.~(\ref{beinvloc}) gives
\begin{displaymath}
\lambda \simeq \frac{6^{1/3} \sqrt{\pi}}{2 a \Gamma \left( 1/6 \right)}
\left( \sigma^{2} \right)^{1/3}
\end{displaymath}
which exhibits the same anomalous scaling found in the Anderson model
at the band-edge~\cite{Der84,Izr98}.
This correspondence is a consequence of the fact that in both models at
the band edge the invariant distribution for the angular variable has
the form~(\ref{berho}) and the ratio $\psi_{n+1}/\psi_{n}$
reduces to the same function of $\theta$.

\section{Existence of anomalously localised states in the Kronig-Penney
model}
\label{stretchexp}

In this section we discuss how specific cross-correlations between the
two kinds of disorder (structural and compositional) can endow the
Kronig-Penney model~(\ref{kpmodel}) with electronic states whose amplitude,
away from the localisation centre $n_{0}$, decays like a stretched
exponential. More precisely, one has
\begin{equation}
|\psi_{n}| \sim \exp \left( - D \sqrt{|n - n_{0}|} \right)
\label{stretchedexp}
\end{equation}
where $D$ is a constant. This corresponds to a stretched exponential
$\exp (-|x|^{\alpha})$ with stretching exponent $\alpha = 1/2$.
The phenomenon has its counterpart in the band-centre anomaly which occurs
in the Anderson model with purely off-diagonal disorder~\cite{Fle77}.

As remarked in the previous section, the Kronig-Penney model~(\ref{kpmodel})
has an equivalent tight-binding model. The correspondence is easily
established by eliminating the momenta from the map~(\ref{hammap1});
in this way, with the obvious substitution $x_{n} \rightarrow
\psi_{n}$, one obtains the equation
\begin{equation}
\begin{array}{ccl}
& & \displaystyle
\frac{1}{\sin \left[ q \left( a + \Delta_{n} \right) \right]} \psi_{n+1} +
\frac{1}{\sin \left[ q \left( a + \Delta_{n-1} \right) \right]} \psi_{n-1} \\
& = & \displaystyle
\left\{ \cot \left[ q \left( a + \Delta_{n} \right) \right] +
\cot \left[ q \left( a + \Delta_{n-1} \right) \right] +
\frac{U + u_{n}}{q} \right\} \psi_{n}.
\end{array}
\label{tbmod}
\end{equation}
It is convenient to express the coefficients of Eq.~(\ref{tbmod})
as sums of their mean values and of fluctuating terms with zero
average. Eq.~(\ref{tbmod}) then assumes the form
\begin{equation}
\left( 1 + \gamma_{n} \right) \psi_{n+1} +
\left( 1 + \gamma_{n-1} \right) \psi_{n-1} +
\varepsilon_{n} \psi_{n} = E \psi_{n}
\label{tibi}
\end{equation}
with $E$ being a deterministic function of the wavevector
$q$ defined by the identity
\begin{equation}
E(q) = \frac{U/q + 2 \Big\langle \cot \left[ q \left( a + \Delta_{n} \right)
\right] \Big\rangle}{\Big\langle 1 / \sin \left[ q \left( a + \Delta_{n} \right)
\right] \Big\rangle} , 
\label{eq}
\end{equation}
while the symbols $\gamma_{n}$ and $\varepsilon_{n}$ stand for the
energy-dependent random variables
\begin{equation}
\gamma_{n}(q) = \frac{1 / \sin \left[ q \left( a + \Delta_{n} \right) \right]}
{\Big\langle 1 / \sin \left[ q \left( a + \Delta_{n} \right) \right]
\Big\rangle} - 1
\label{gamvar}
\end{equation}
and
\begin{equation}
\begin{array}{ccl}
\varepsilon_{n}(q) & = & \displaystyle
\frac{1}{\Big\langle 1 / \sin \left[ q \left( a +
\Delta_{n} \right) \right] \Big\rangle}
\left\{ 2 \Big\langle \cot \left[ q \left( a + \Delta_{n} \right)
\right] \Big\rangle - \cot \left[ q \left( a + \Delta_{n} \right) \right]
\right. \\
& - & \displaystyle \left.
\cot \left[ q \left( a + \Delta_{n-1} \right) \right] - \frac{u_{n}}{q}
\right\} .
\end{array}
\label{epsvar}
\end{equation}
Eq.~(\ref{tibi}) shows that the tight-binding counterpart of the
Kronig-Penney model~(\ref{kpmodel}) is an Anderson model with both
diagonal and off-diagonal disorder.
Note that, in the absence of structural disorder, the random
variables~(\ref{gamvar}) vanish and the Kronig-Penney's analogue becomes
the ordinary Anderson model with diagonal disorder. For purely compositional
disorder, therefore, one could have predicted {\em a priori} the existence
of the anomalies discussed in Sec.~\ref{anomal}.

We now focus our attention on the case in which the compositional disorder
has the form
\begin{equation}
u_{n} = 2q_{c} \Big\langle \cot \left[ q_{c} \left( a + \Delta_{n} \right)
\right] \Big\rangle -
q_{c} \cot \left[q_{c} \left( a + \Delta_{n} \right) \right] -
q_{c} \cot \left[q_{c} \left( a + \Delta_{n-1} \right) \right]
\label{zerodiagdis}
\end{equation}
where $q_{\rm c}$ represents the wavevector fulfilling the condition
\begin{equation}
E(q_{c}) = 0 .
\label{zeroenergy}
\end{equation}
We stress that condition~(\ref{zerodiagdis}) introduces special
cross-correlations between the two kinds of disorder in the Kronig-Penney
model~(\ref{kpmodel}).
We also remark that the weak-disorder condition~(\ref{weaksd}) ensures
that a solution of Eq.~(\ref{zeroenergy}) exists. In fact, taking into
account the band-structure relation~(\ref{kpbandstructure}), within
the limits of the second-order approximation one can cast
Eq.~(\ref{zeroenergy}) in the form
\begin{displaymath}
\cos \left( k_{\rm c} a \right) \simeq
\frac{q_{\rm c} \langle \Delta_{n}^{2} \rangle U}{2 \sin (q_{\rm c} a)} .
\end{displaymath}
If the structural disorder is weak enough, the right-hand side of this
equation is less than one and this implies that a Bloch vector $k_{\rm c}$
exists such that $q(k_{\rm c})$ is the solution of Eq.~(\ref{zeroenergy}).
Actually, a perturbative calculation shows that the Bloch vector $k_{\rm c}$
in the positive half of the first Brillouin zone is
\begin{equation}
k_{\rm c} \simeq
\frac{\pi}{2a} - \frac{\overline{q} \langle \Delta_{n}^{2} \rangle U}
{2 a \sin (\overline{q}a)}
\label{kcrit}
\end{equation}
with $\overline{q} = q \left( \frac{\pi}{2a} \right)$.
For weak disorder the deviation of $k_{\rm c}$ from $\pi/2a$ is not large
and therefore $q_{\rm c}^{2} \simeq {\overline{q}}^{2}$, which is
close to centre of the energy band.

If the compositional disorder has the form~(\ref{zerodiagdis}), it
is easy to see that, when the electron energy takes the critical value
$q_{c}^{2}$ identified by the condition~(\ref{zeroenergy}),
Eq.~(\ref{tibi}) becomes
\begin{equation}
\left[ 1 + \gamma_{n}(q_{\rm c}) \right] \psi_{n+1} +
\left[ 1 + \gamma_{n-1}(q_{\rm c}) \right] \psi_{n-1} = 0 ,
\label{offdiaganderson}
\end{equation}
which has the same form of the Schr\"{o}dinger equation for the Anderson
model with purely off-diagonal disorder and zero energy.
For zero energy, the latter model is known to have an electronic state 
which exhibits anomalous localisation, because it is localised but decays
away from the localisation centre $n_{0}$ according to
Eq.~(\ref{stretchedexp})~\cite{Fle77}.
We stress that the identity of Eq.~(\ref{offdiaganderson}) with the
zero-energy Schr\"{o}dinger equation for the Anderson model with only
off-diagonal disorder ensures that, when the compositional
disorder has the form~(\ref{zerodiagdis}) and the energy takes the critical
value $q_{c}^{2}$, the Kronig-Penney model also has an anomalously localised
state whose amplitude decays exponentially with the square root of the
distance from the localisation centre.

This property can be heuristically justified with the observation that
Eq.~(\ref{offdiaganderson}) implies that
\begin{equation}
\log |\psi_{2n}| = \log |\psi_{0}| +
\sum_{l=0}^{n-1} \left[ \log \left| 1 + \gamma_{2l}(q_{\rm c}) \right|
- \log \left| 1 + \gamma_{2l+1}(q_{\rm c}) \right| \right]
\label{sumrv}
\end{equation}
By invoking the central limit theorem, one can therefore conclude that,
for large values of $n$, the random variable $\log |\psi_{n}|$ has
zero average and a variance which increases linearly with $n$.
We stress that the previous argument holds even if the disorder
exhibits long-range self-correlations of the form
\begin{equation}
\chi(l) = \frac{1}{c_{2} - c_{1}}\frac{1}{\pi l}
\left[ \sin \left( \pi c_{2} l \right) -
\sin \left( \pi c_{1} l \right) \right]
\label{lrbincor}
\end{equation}
where $c_{1}$ and $c_{2}$ are real numbers such that $0 < c_{1} < c_{2}
\leq 1$.
In fact, weaker forms of the central limit theorem can be applied to sums
of correlated random variables, provided that the correlations decay fast
enough~\cite{Ma85,Dav89}.
Specifically, given a succession of zero-average, correlated random
variables $\{ x_{n} \}$, let $S_{N} = \sum_{n=1}^{N} x_{n}$ be the sum of
the first $N$ terms of the succession.
The minimal condition for the mean square of $S_{N}$ to grow linearly with
$N$, i.e., $\langle \left( S_{N} \right)^{2} \rangle \sim N$, is that
the power spectrum of the succession $\{ x_{n} \}$ be finite at the origin.
The power spectrum corresponding to the binary correlator~(\ref{lrbincor})
is
\begin{equation}
W(ka) = \left\{ \begin{array}{ccc}
\displaystyle
\frac{1}{c_{2} - c_{1}} &
\mbox{ if } & \displaystyle
ka \in \left[ c_{1} \frac{\pi}{2}, c_{2} \frac{\pi}{2} \right] \cup
                     \left[ \pi - c_{2} \frac{\pi}{2},
                     \pi - c_{1} \frac{\pi}{2} \right] \\
0 & & \mbox{ otherwise }
\end{array} \right.
\label{powspe}
\end{equation}
and vanishes at the origin; one can therefore conclude from Eq.~(\ref{sumrv})
that
\begin{displaymath}
\log |\psi_{n}| \sim \sqrt{n}
\end{displaymath}
even if the disorder is correlated.

\begin{figure}[t]
\begin{center}
\epsfig{file=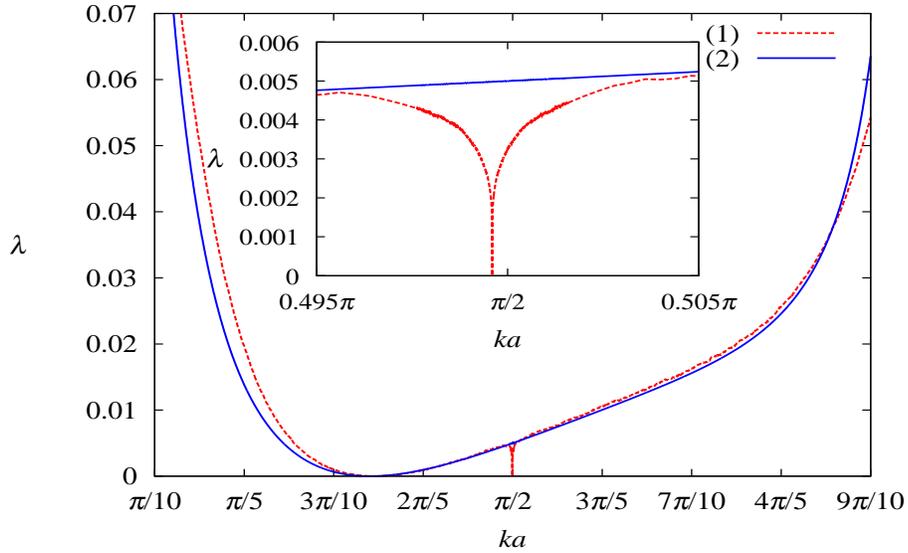,width=5in,height=3in}
\caption{(Colour on line)
Inverse localisation length $\lambda$ versus $k a$ for the case
of structural disorder without self-correlations.
The dashed line (1) represents numerical data, while the solid line (2)
corresponds to Eq.~(\ref{thlyap}). The inset shows the anomaly in greater
detail.
\label{anoloc}}
\end{center}
\end{figure}

\begin{figure}[t]
\begin{center}
\epsfig{file=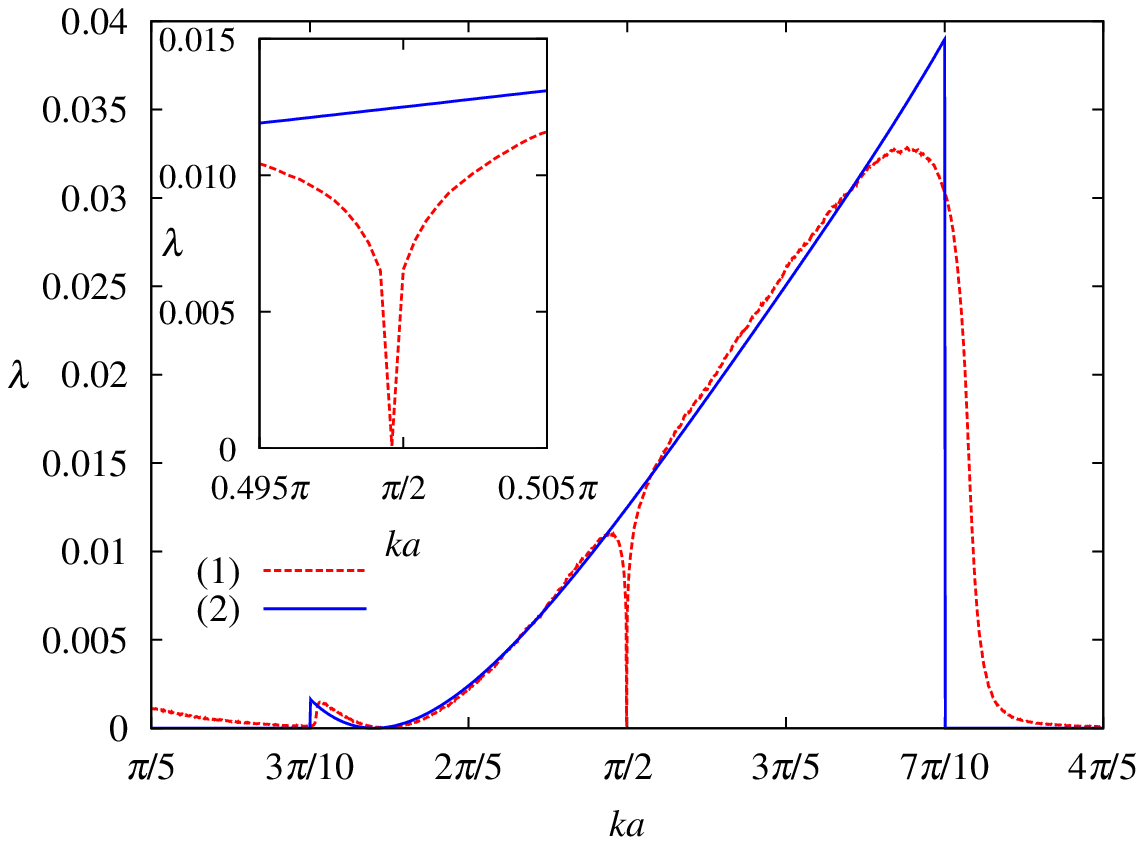,width=5in,height=3in}
\caption{(Colour on line)
Inverse localisation length $\lambda$ versus $k a$ for the
case of self-correlated structural disorder.
The dashed line (1) represents numerical data, while the solid line (2)
corresponds to Eq.~(\ref{thlyap}). The inset shows the anomaly in greater
detail.
\label{anoloc_cor}}
\end{center}
\end{figure}

The numerical data confirm the conclusion that the Kronig-Penney
model~(\ref{kpmodel}) has an anomalously localised state for $q = q_{\rm c}$
when the compositional disorder takes the special form~(\ref{zerodiagdis}).
This can be seen from Figs.~\ref{anoloc} and~\ref{anoloc_cor}, which show
how the Lyapunov exponent vanishes when the Bloch wavevector takes the
critical value $k_{\rm c}$.
The difference between Figs.~\ref{anoloc} and~\ref{anoloc_cor} lies in the
fact that in the case corresponding to Fig.~~\ref{anoloc} the structural
disorder is not self-correlated, while the data represented in
Fig.~\ref{anoloc_cor} were obtained for structural disorder with long-range
self-correlations of the form~(\ref{lrbincor}) with $c_{1} = 3/10$ and
$c_{2} = 1$.

All the numerical data presented in this section were obtained
for a Kronig-Penney model with mean field $U = 4$ and structural disorder
characterised by a uniform distribution
\begin{equation}
p(\Delta_{n}) = \left\{ \begin{array}{ccl}
1/W & & \mbox{ if } \Delta_{n} \in [-W/2, W/2] \\
0 & & \mbox{ otherwise}
\end{array} \right.
\label{boxdis}
\end{equation}
with width $W = 0.1732\ldots$, corresponding to a disorder strength
$\sqrt{\langle \Delta_{n}^{2} \rangle} = 0.05$.
For the above-specified values of the mean field and of the disorder
strength, formula~(\ref{kcrit}) gives a value of the critical Bloch
wavevector approximately equal to $k_{\rm c} \simeq 0.4952 \pi/a$, in
relatively good agreement with the numerically obtained value
$k_{\rm c} \simeq 0.4996 \pi/a$.

When the box distribution~(\ref{boxdis}) is chosen for the displacements
$\Delta_{n}$, the function~(\ref{eq}) becomes
\begin{equation}
E(q) = \frac{\displaystyle UW + 2 \log \frac{\sin \left[ q \left(
a + W/2 \right) \right]}{\sin \left[ q \left( a - W/2 \right) \right]}}
{\displaystyle \log \frac{\sin (qa) + \sin (qW/2)}
{\sin (qa) - \sin (qW/2)}} ,
\label{eboxdis}
\end{equation}
while the random variables~(\ref{gamvar}) and~(\ref{epsvar}) take
the forms
\begin{equation}
\gamma_{n}(q) = \frac{1}{\sin \left[ q \left( a + \Delta_{n} \right) \right]}
\frac{qW}{\displaystyle \log \frac{\sin (qa) + \sin (qW/2)}
{\sin (qa) - \sin (qW/2)}}
\label{gammaboxdis}
\end{equation}
and
\begin{equation}
\begin{array}{ccl}
\varepsilon_{n}(q) & = & \displaystyle
\frac{1}{\displaystyle \log \frac{\sin (qa) + \sin (qW/2)}
{\sin (qa) - \sin (qW/2)}} \Bigg\{
2 \log \frac{\sin \left[ q \left( a + W/2 \right) \right]}
{\sin \left[ q \left( a - W/2 \right) \right]} \\
& - & \displaystyle
qW \cot \left[ q \left( a+ \Delta_{n} \right) \right] -
qW \cot \left[ q \left( a+ \Delta_{n-1} \right) \right] -
W u_{n} \Bigg\} .
\end{array}
\label{epsboxdis}
\end{equation}
We would like to stress that, when performing numerical calculations, one
should work with the exact form~(\ref{eboxdis}) of $E(q)$ and the exact
expressions~(\ref{gammaboxdis}) and~(\ref{epsboxdis}) of the
coefficients $\gamma_{n}$ and $\varepsilon_{n}$, {\em even if the
disorder is weak}. Obviously, the explicit expressions of these magnitudes
depend on the distribution chosen for the variables $\Delta_{n}$ and
must be modified if the box distribution is replaced with another one.
Whatever distribution is adopted, however, it is important that the
corresponding exact expressions of $E(q)$, $\gamma_{n}$ and
$\varepsilon_{n}$ be used.
Using second-order approximations for $E(q)$ and the random
coefficients~(\ref{epsvar}) works relatively well for most values of
the energy, but fails at the critical point, because the neglected
higher-order corrections produce non-zero diagonal terms in
Eq.~(\ref{tibi}) which, in spite of being very small, prevent the
electronic state from being anomalously localised and the Lyapunov
exponent from vanishing completely.

As a side remark, we would like to add that the
conditions~(\ref{zerodiagdis}) and~(\ref{zeroenergy}) ensure the
existence of an anomalously localised state also for disorder of
arbitrary strength, provided that Eq.~(\ref{zeroenergy}) has a
solution $q_{c}$ inside the allowed energy bands. This is confirmed
by numerical calculations (which, incidentally, also show that when
disorder is not weak the critical value of the energy need not be
close to the band centre).
To determine the conditions which guarantee the existence of such
a critical value of the energy in the general case is not an easy task,
however; for the sake of simplicity, we therefore restrict our attention
to the case of weak disorder, for which the band structure is
approximately given by Eq.~(\ref{kpbandstructure}) and Eq.~(\ref{zeroenergy})
does have a solution.

In both Fig.~\ref{anoloc} and~\ref{anoloc_cor} the numerical data are
compared with the theoretical predictions derived from the general
result~(\ref{invloc0}). Because in this formula terms of order higher
than the second are neglected, in its evaluation we replaced
the exact expression~(\ref{zerodiagdis}) of the compositional disorder
with its second-order approximation
\begin{equation}
u_{n} \simeq \frac{q_{\rm c}^{2}}{\sin^{2}(q_{\rm c} a)}
\left( \Delta_{n} + \Delta_{n-1} \right) .
\label{approxzerodiagdis}
\end{equation}
In passing, we observe that Eq.~(\ref{approxzerodiagdis}) implies that
the cross-correlator $\chi_{3}(l)$ is {\em not} an even function of its
argument.
When the compositional disorder has the form~(\ref{approxzerodiagdis}),
the inverse localisation length~(\ref{invloc0}) becomes
\begin{equation}
\lambda = \frac{1}{8a} \langle \tilde{\Delta}_{n}^{2} \rangle
\left[ 1 - 2 \frac{q_{\rm c}^{2} \sin (q a)}{U q \sin^{2} (q_{\rm c} a)}
\cos (ka) \right]^{2} W_{2}(ka) .
\label{thlyap}
\end{equation}
The power spectrum $W_{2}(ka)$ in the previous formula reduces to
$W_{2}(ka) = 1$ in the case of uncorrelated structural disorder,
while it is of the form~(\ref{powspe}) in the case of self-correlated
structural disorder represented in Fig.~\ref{anoloc_cor}.
Note that in the latter case the long-range correlations~(\ref{lrbincor})
create two mobility edges at $ka = 3 \pi/10$ and $ka = 7 \pi/10$.
As can be seen from Figs.~\ref{anoloc} and~\ref{anoloc_cor}, the theoretical
formula~(\ref{thlyap}) works reasonably well everywhere, except in a
small neighbourhood of the critical value $k_{\rm c}$.
This failure must be ascribed to the fact that Eq.~(\ref{thlyap}), as its
parent expression~(\ref{invloc0}), is not valid for Bloch wavevectors lying
close to the rational value $\pi/2a$, where resonance effects play a
non-negligible role.

As a last comment on the specific features of the localisation length in
the special case in which the compositional disorder is related to the
structural one by Eq.~(\ref{zerodiagdis}), we observe that Eq.~(\ref{thlyap})
predicts the existence of a delocalised state for the Bloch wavevector
$k^{\star}$ identified by the condition
\begin{displaymath}
2 \frac{q_{\rm c}^{2} \sin \left( q(k^{\star})a \right)}{q(k^{\star})
\sin^{2} \left( q_{\rm c} a \right)} \cos \left( k^{\star} a \right) = U .
\end{displaymath}
This is confirmed by the numerical computations, as can be seen from
both Figs.~\ref{anoloc} and~\ref{anoloc_cor}.

\begin{figure}[t]
\begin{center}
\epsfig{file=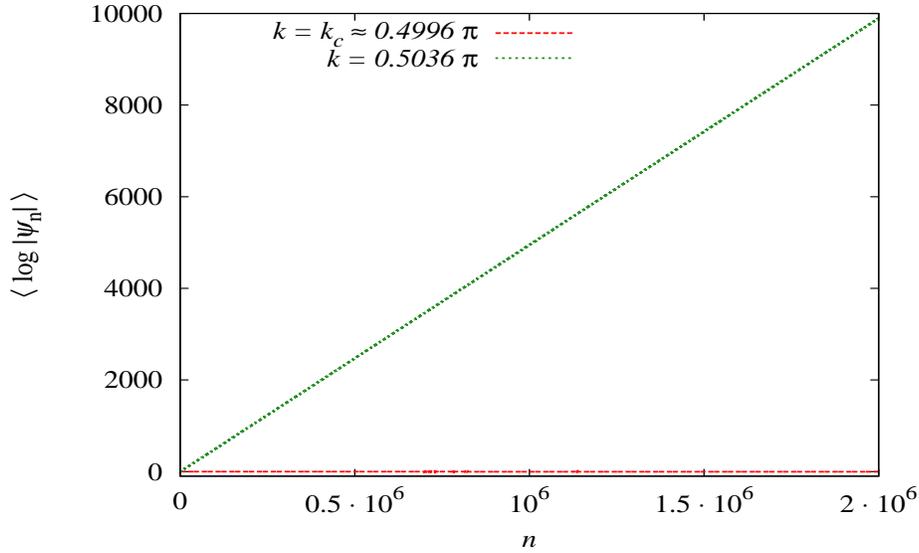,width=5in,height=3in}
\caption{(Colour on line)
$\langle \log |\psi_{n}| \rangle$ versus $n$. The dashed line corresponds
to the critical value of the Bloch wavevector, $k = k_{\rm c} \simeq 0.4996
\pi$, while the dotted line corresponds to a Bloch wavevector $k = 0.5036 \pi$.
\label{m1}}
\end{center}
\end{figure}

\begin{figure}[t]
\begin{center}
\epsfig{file=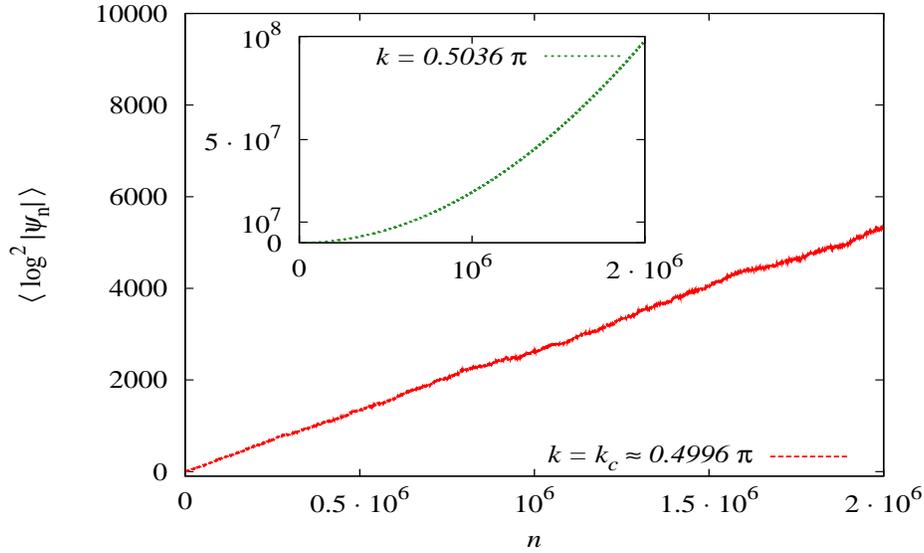,width=5in,height=3in}
\caption{(Colour on line)
$\langle \log^{2} |\psi_{n}| \rangle$ versus $n$. The dashed line corresponds
to the critical value of the Bloch wavevector, $k = k_{\rm c} \simeq 0.4996 \pi$,
while the dotted line in the inset corresponds to a Bloch wavevector
$k = 0.5036 \pi$.
\label{m2}}
\end{center}
\end{figure}

To conclude this section, we would like to add some numerical evidence
of the stretched-exponential behaviour of the tails of the anomalously
localised state. We have numerically solved Eqs.~(\ref{tibi})
and~(\ref{offdiaganderson}) as initial-value problems; this corresponds to
constructing the electronic states with the transfer-matrix technique.
In the anomalous case one expects $\log |\psi_{n}|$ to behave as the
position of a random walker, i.e., as a random variable with constant
zero average and a second moment linearly increasing with $n$.
For $k \neq k_{\rm c}$, on the other hand, the solution of Eq.~(\ref{tibi})
should behave as $|\psi_{n}| \sim \exp (\lambda n)$, leading to an increase
with $n^{l}$ of the $l$-th moment of $\log |\psi_{n}|$.
This is confirmed by the numerical data for the first two moments
of the variable $\log |\psi_{n}|$, represented in Figs.~\ref{m1}
and~\ref{m2}.
In both Figs.~\ref{m1} and~\ref{m2} we considered the behaviour of
$\log |\psi_{n}|$ as a function of $n$ for two Bloch wavevectors, i.e.,
the critical vector $k = k_{\rm c} \simeq 0.4996 \pi$ and the vector
$k = 0.5036 \pi$, which is close to the critical value but not identical
to it.
The moments of $\log |\psi_{n}|$ were computed with an average over 1000
disorder realisations. For the sake of simplicity, we considered
compositional disorder without self-correlations.
Both the first and the second moment of $\log |\psi_{n}|$ behave as
expected, corroborating the conclusion that the tails of the electronic
state at the critical point are described by Eq.~(\ref{stretchedexp}).

\section{Conclusions}
\label{conclu}

In this work we analyse the anomalous behaviour of specific electronic
states in the Kronig-Penney model with weak compositional and structural
disorder. In every case we discuss the analogies with the corresponding
phenomena in the Anderson model.

We first show that the localisation length deviates from the prediction
of the standard formula~(\ref{invloc0}) when the Bloch vector assumes the
value $k = \pi/2a$, which corresponds to an energy close to the band
centre. This discrepancy is due to the same resonance effect which occurs
in the standard Anderson model at the band centre; in both models this
effect produces a modulation of period $\pi/2$ of the invariant distribution
of the angle variable of the associated Hamiltonian map. This modulation
leads to the band-centre anomaly of the localisation length.

We make use of the Hamiltonian map approach also to analyse the
electronic states at the band edge. We find again that the same anomalous
behaviour originally found in the Anderson model is present in the
Kronig-Penney model.
In both systems the most relevant feature at the band edge is the
anomalous scaling of the localisation length with the disorder strength.

We finally use the correspondence between the Kronig-Penney model and the
Anderson model with diagonal and off-diagonal disorder to conclude that
in the former system specific cross-correlations between the two kinds of
disorder generate an electronic state, close to the band centre, whose
tails decay as stretched exponentials.
This state is the analogue of the anomalously localised state which
occurs at the band-centre in the Anderson model with purely off-diagonal
disorder.

J. C. H.-H. and L.T. gratefully acknowledge the support of the CONACyT grant
No. 84604 and of the CIC-2009 grant (Universidad Michoacana).
The work of F.M.I. was partly supported by the CONACyT grant No. 80715.

\end{document}